\documentstyle[11pt,newpasp,twoside,epsf]{article}

\newcounter{subfigure}

\begin{document}

\title{ A Large Extension of the CfA Galactic CO Survey}

\author{T. M. Dame and P.  Thaddeus}

\affil{Harvard-Smithsonian Center for Astrophysics, 60 Garden Street, Cambridge, MA 02138, USA}

\begin{abstract} The Galactic CO survey of Dame, Hartmann, \& Thaddeus (2001; 
hereafter DHT) is composed of both large-scale unbiased surveys, mainly concentrated 
within 10{\deg} of the Galactic plane, and targeted observations of clouds at higher 
latitudes.  Analysis of all-sky IRAS and 21 cm maps suggests that the DHT survey is nearly 
complete for clouds larger than $\sim1${\deg}, even though roughly half of the total area 
at $|b| < 30^\circ$ was not observed.  In October 2001 we began a new survey of all of 
this unobserved area that is accessible to the northern 1.2 meter telescope, 
approximately 6,600 deg\(^{2}\) between \(l\) = 0{\deg} and 230{\deg}, mainly in the latitude range $|b|$ = 10{\deg}--30{\deg}. At least 12 hours per day is being dedicated to this large project, which is sampled every  {\onequarter}{\deg} (every other beamwidth) to an rms sensitivity of 0.19 K at a velocity resolution of 0.65 {km~s$^{-1}$}.  As of May 2003, we have obtained 90,000 of the 106,000 spectra required to complete the survey. While, as expected, the new observations do not substantially change the  DHT map, ~68 relatively small and isolated molecular clouds at intermediate latitudes have so far been discovered. 

The survey has also been extended to $b < -30^\circ$ in two regions, one in the vicinity of
the large MBM clouds 53-55 at $l \sim90${\deg} and the other south of the Taurus clouds at
$l \sim170${\deg}. Substantial amounts of molecular gas were detected in both of these 
high-latitude regions.
\end{abstract}

\section{Introduction}

The 1.2 meter millimeter-wave telescopes at the Harvard-Smithsonian 
Center for Astrophysics and at CTIO in Chile have been playing a leading role 
in surveying the molecular gas in our Galaxy for over two decades.  A 
significant milestone of this project was reached two years ago, 
with the publication of the composite CO survey of Dame, Hartmann, \& 
Thaddeus (2001; hereafter DHT). The DHT survey has 16 times more spectra 
than the previous composite
survey at {\onehalf}{\deg} angular resolution done with the same telescopes (Dame et al. 1987),
up to 3.4 times higher angular resolution, and up to 10 
times higher sensitivity per unit solid angle. 

In DHT it was argued that the survey was largely 
complete at $|b| < 30^\circ$, even though roughly half of that area had not been 
surveyed.  For several years prior to the publication of DHT, we had used the 
IRAS far-infrared survey as a tracer of total gas, and the Leiden-Dwingeloo 21 
cm survey as a tracer of atomic gas, in order to deduce from the differences between the
two the existence of molecular clouds outside the limits of our large Galactic plane surveys;
about a dozen clouds were found in this way and subsequently mapped in CO. 
DHT used the IRAS 
and Leiden-Dwingeloo 21 cm surveys to show that their molecular survey was nearly 
complete for clouds brighter in CO than $\sim1$ K km s\(^{-1}\) deg\(^{2}\); for a typical high-latitude cloud at a distance of 100 pc, this brightness corresponds to an H\(_{2}\) 
mass of only 9 M\(_{\sun}\) and an angular size of about 1{\deg}.  Smaller clouds could slip
through the net of the analysis, since it was limited by the $36\arcmin$  resolution of the 21 cm 
survey and by small-scale variations in the dust temperature and gas-to-dust 
ratio.   

Here we describe our efforts over the past three years to fill 
in all of the DHT survey area accessible to our northern telescope with 
uniform sampling every {\onequarter}{\deg}.  
In terms of number of spectra and total integration time, 
this is by far the largest single survey 
undertaken with either of the 1.2 m telescopes, requiring up to 18 hours per day 
for three 7-month observing seasons.  
In the next section we will describe the 
observations in detail, then go on to discuss some preliminary results from 
the survey, which is now about 85\% complete.

\renewcommand{\thefigure}{\arabic{figure}\alph{subfigure}}
\setcounter{subfigure}{1}
\begin{figure} \plotone{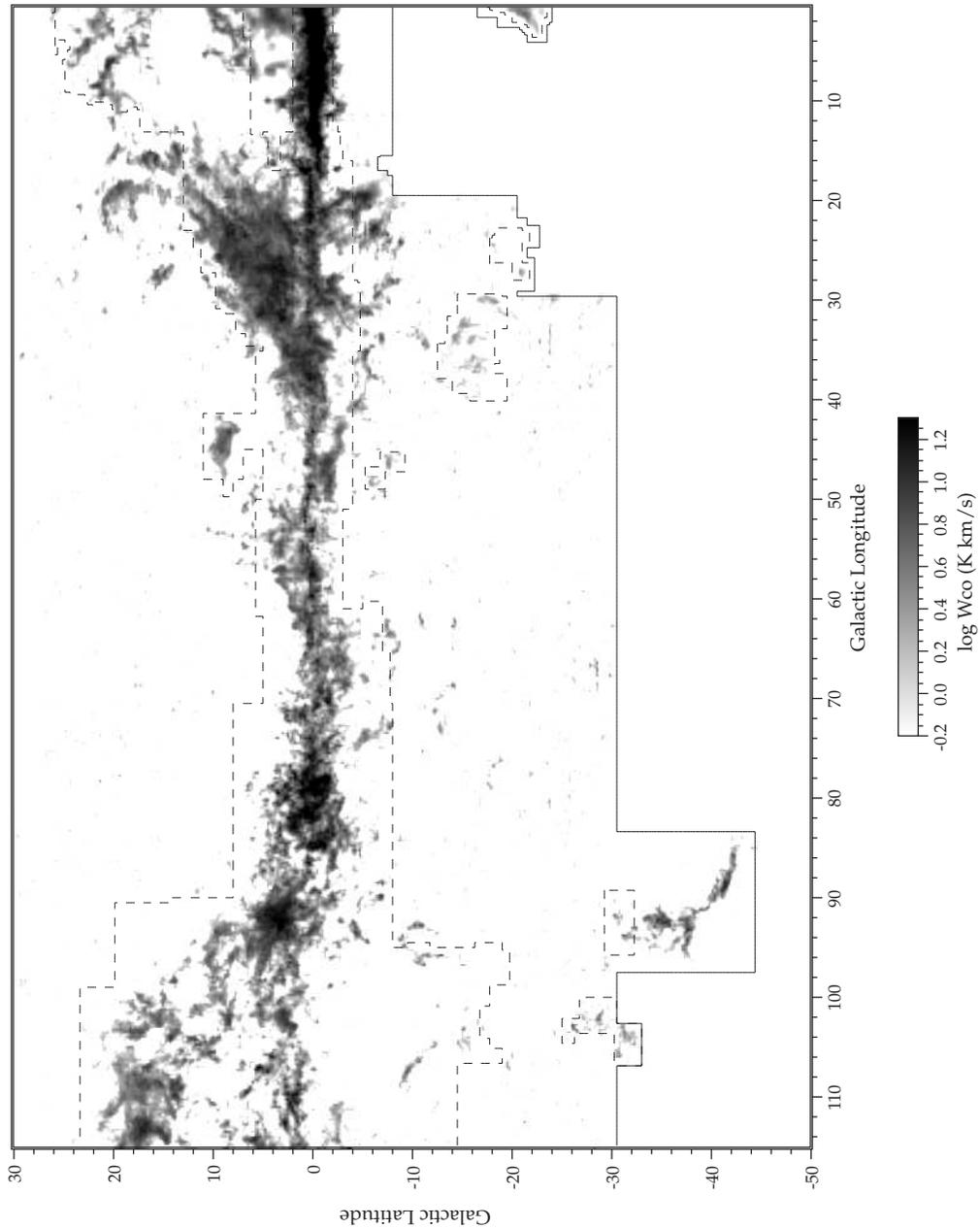} \caption{ A velocity-integrated CO map which combines
DHT observations with better than {\onehalf}{\deg} resolution (within the dashed line) with 
the new survey here (between the dashed and solid lines). Only half of the new survey area is shown here; the other half is shown in Figure 1{\it b}. Since the new survey is still on-going, this
map should be considered preliminary. } \end{figure}

\addtocounter{figure}{-1}
\addtocounter{subfigure}{1}
\begin{figure} \plotone{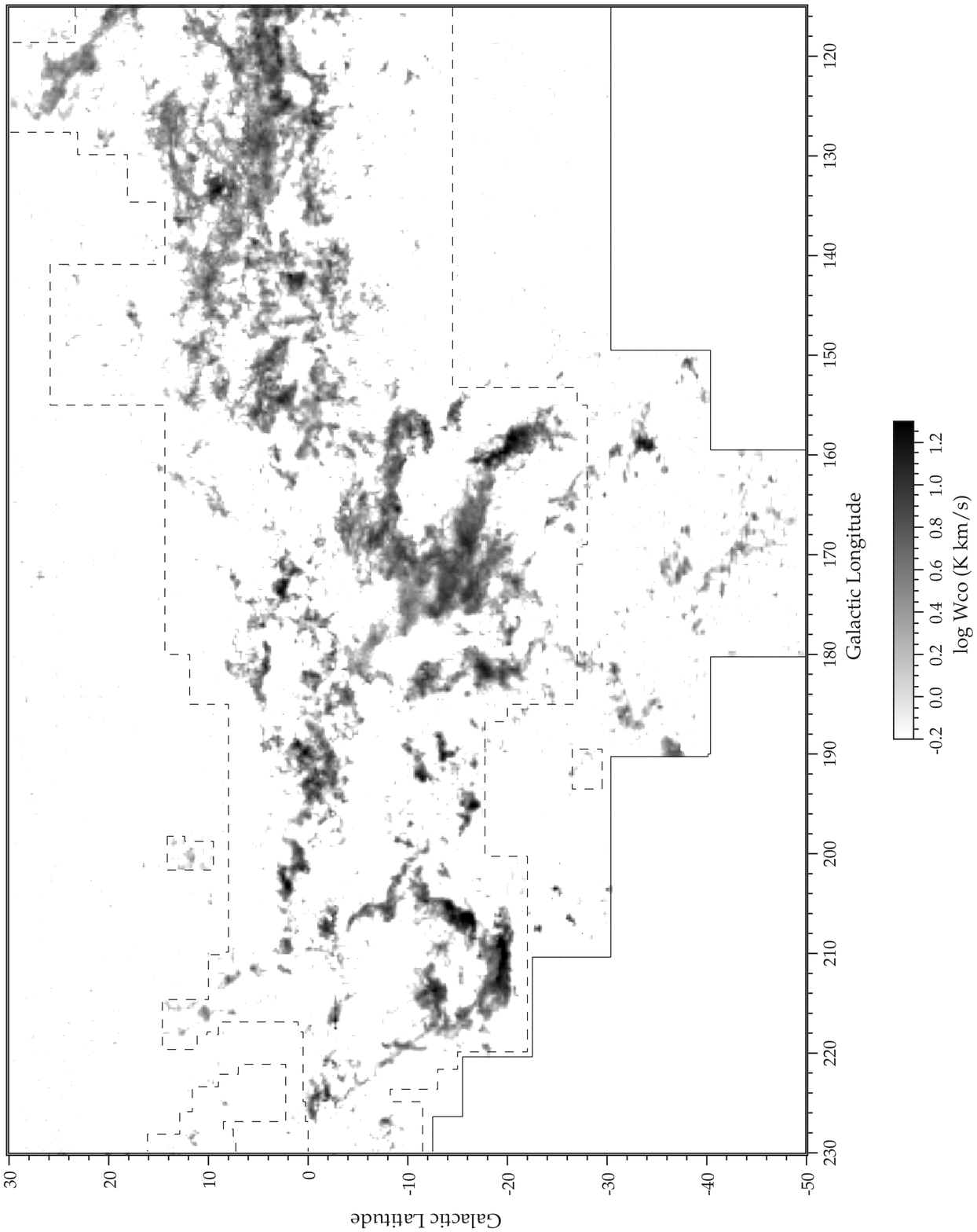} \caption{ A continuation of Figure 1a through the second
Galactic quadrant and beyond. } \end{figure}

\renewcommand{\thefigure}{\arabic{figure}}

\section{Observations}

Our new CO survey includes all of the area at $|b| < 30^\circ$
and $\delta > -17^\circ$ that is not covered by the DHT survey, 
or covered only by survey No. 1 in that paper, which has an effective angular resolution of only {\onehalf}{\deg} (Dame et al. 1987); the total survey area is 6600 deg\(^{2}\).
 The survey sensitivity of 0.19 K in 0.65 {km~s$^{-1}$} spectral channels is typical of our
Galactic plane surveys, and is adequate to detect molecular clouds with visual extinctions as low as 0.1 mag.  The survey is being conducted by sequentially observing 4 interleaved 
{\onehalf}{\deg} grids, for a final sampling interval of {\onequarter}{\deg}---slightly better than every other beamwidth ($8.4\arcmin$). After 3 full observing seasons we have obtained 90,000 of the 106,000 spectra required to complete the survey at $|b| < 30^\circ$. Recently we have also extended the survey to two regions at $b < -30^\circ$, one near the high-latitude cloud complex MBM 53-55, and the other south of the Taurus dark clouds.

The telluric emission line from CO in the mesosphere is potentially a serious
source of contamination in the present survey, 
which is being done with frequency switching every 1 second over 15 MHz---
an efficient mode of data acquisition with the liability that it 
does not chop out a widely distributed telluric line. 
When mapping a particular region it is often possible, by observing at 
the right time, to position the telluric line in LSR velocity well away from the 
celestial CO emission. In the present survey, however, observations are 
conducted over the whole sky every day, and the emission observed is often 
weak and at a velocity which is not precisely known in advance. The telluric line 
removal therefore has to be handled carefully. A model of the telluric 
line is fit to all spectra taken on a given day, or in some shorter interval if the 
line intensity varies rapidly, as it sometimes does. We eliminate 
from the fit any telluric lines which appear to overlap celestial emission. This 
model is then subtracted from all spectra taken during that time interval, for
cancellation of the telluric line to below the noise in individual spectra 
and to better than 2\% on average. 
(A useful byproduct of the present survey is high signal-to-noise daily 
monitoring of the telluric line parameters over three years. Besides well-known 
seasonal variations, we occasionally detect variations in the telluric line 
intensity by more than 50\% in less than a day.)

\section{ Results at $|b| < 30${\deg}}

The molecular Galaxy over the full longitude range of the present survey is shown in 
Figures 1{\it a} and 1{\it b}. The area mapped previously by DHT at better 
than {\onehalf}{\deg}
resolution is enclosed by the dashed line, and the limits of the new survey are marked by 
the solid line. It is to be emphasized that the present survey is still underway,
and some regions are nosier than others owing to incomplete sampling.
The observations at $b < -30^\circ$ will be discussed in the next section.  
Here we focus on the region of the DHT map, $|b| < 30^\circ$.  
As predicted by DHT, only small clouds were found in the previously unobserved regions. 
The large foldout map published by DHT is therefore a remarkably complete inventory of
molecular clouds in the Galaxy within $30^\circ$ of the plane. 
One exception is the region of
the Chamaeleon clouds, which lies in the southern sky beyond the reach of the 
present survey.  In DHT Chamaeleon was identified as the one region where 
substantial CO emission might have been missed, since the sampling there was 
tightly confined to the three main clouds. Observations by the NANTEN 
telescope reported at this meeting (Mizuno et al.) do indeed show a substantial 
amount of CO emission in that region.  

Although the new observations here add little to the DHT 
map, they have revealed about 68 small, isolated clouds or 
clouds clusters, most barely visible on the scale of Figure 1, that are apparently 
unrelated to large well-known molecular objects such as the Polaris Flare or the Aquila Rift. 
There are in addition a few dozen more tentative detections 
which require confirmation. 
Since only a few of the 68 definite detections match the high-latitude 
clouds cataloged by Magnani, Blitz, \& Mundy (1985; hereafter MBM), 
it is likely that most were entirely unknown before the present work. 
Before finishing the survey, we will 
obtain fully-sampled maps of all these objects to better determine masses and
internal structures. These high-latitude objects are presumably quite close,
and may be the closest molecular objects in the Galaxy; they may be useful
in tracing out large-scale spatial and kinematic structures such as supershells. 

Even with the new clouds here, the region at $|b|$ = 20{\deg}--30{\deg} is so deficient in molecular gas that it is reasonable to ask 
whether our detection rate there is less than expected on extrapolating from the distribution in
the plane. On the assumption that we are located near the center of a 
plane-parallel uniform distribution of molecular gas with a mean surface density of 0.87 M\(_{\sun}\) pc\(^{-2}\) and a thickness of 87 pc (Dame et al. 1987), we calculate that the
observed detection rate is indeed about a factor of 5 smaller than expected. 
Because much of the local molecular gas, however, is not uniformly distributed, but is 
concentrated in a few large clouds like those in Taurus, this deficiency is plausibly
a statistical fluctuation. 

\section{The Taurus-South and MBM 54 Extensions}

Several years ago Magnani et al. (2000) used the Northern 1.2 m telescope 
to survey the entire southern Galactic hemisphere at $b < -30^\circ$ and 
$\delta > -17^\circ$ every one degree.  While the overall filling factor of CO in that region 
was found to be low (0.03), two regions showed a fairly large number of detections, 
one south of the Taurus clouds (hereafter, the Taurus-S region), the other 
in the vicinity of the MBM 53, 54, and 55 clouds. In the past year we have 
extended our survey to cover both of these regions, and as Figure 1 shows, 
both as suspected contain substantial amounts of molecular gas. 

Our mapping of the MBM 53-55 clouds, near $l \sim90${\deg}, was completed by 
E. Aguti, who has a separate contribution in this volume on these interesting 
objects.  It was not clear from the sparse mapping in MBM that the individual 
clouds MBM 53, 54, and 55 were related, but Aguti's more sensitive and 
uniformly-sampled map shows that almost certainly they are, since they all 
lie within a common envelop of emission in a region otherwise nearly 
devoid of molecular gas.  Further, as Aguti et al. show, all three clouds lie 
along the edge of an apparent expanding shell of H I (noted also by Gir et al. 
1994). Onishi also has a contribution here describing new \(^{12}\)CO, 
\(^{13}\)CO and C\(^{18}\)O observations of MBM 53-55 with the NANTEN telescope.  

\begin{figure} \plotone{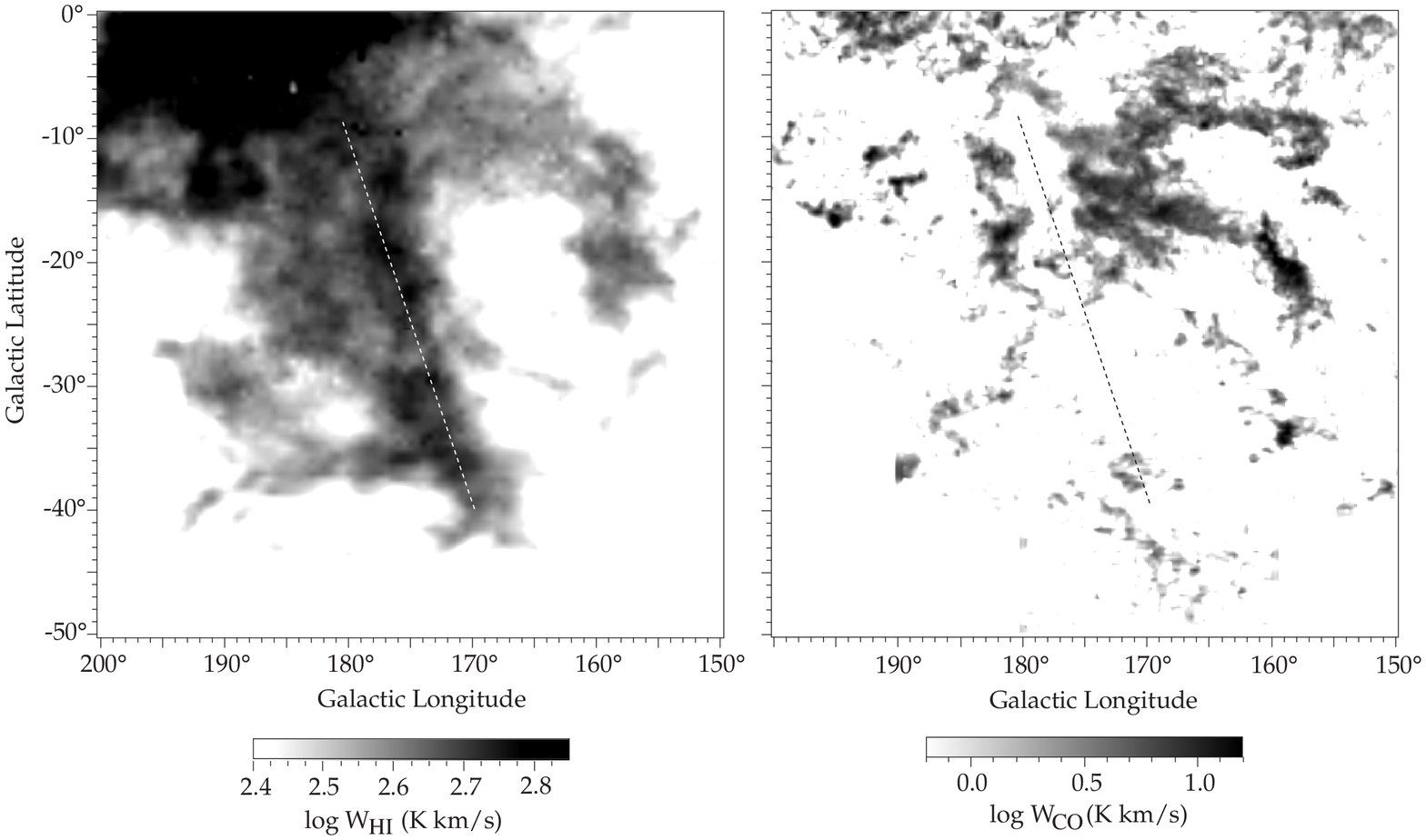} \caption{{\it Left:} 21 cm emission from the
survey of Hartmann \& Burton (1997), integrated from 2 to 10 km s\(^{-1}\). The
grayscale spans a relatively small range of intensities in order to emphasize the
worm-like feature. {\it Right:} Velocity-integrated CO intensity, as in Figure 1.
The region at $b < -30^\circ$ is still irregularly sampled at slightly better than
{\onehalf}{\deg}. The same sloping dashed line is plotted in both figures to 
aid comparison.} \end{figure}

In the Taurus-S region, less than three of the four projected half-degree grids are 
now complete.  It is however already clear that for a region so far from the plane, 
Taurus-S contains a substantial amount of molecular gas.  Because of the rarity of molecular gas 
more than 10{\deg} from the plane in Figure 1, it seems likely that the Taurus-S 
emission is related to the main Taurus-Auriga cloud complex. Magnani 
(1988) has similarly argued, from the distribution of IRAS emission in the region
and on limited distance information, that all 8 MBM clouds 
in this region are probably related to the Taurus clouds. 

In Taurus-S, one lane of clouds seems to extend down and to higher longitude 
from a vertical concentration of clouds at the edge of the 
Taurus clouds (near $l$ = 182{\deg}), while another extends down and to lower 
longitude from the vicinity of the Pleiades ($l$ = 166.5{\deg}, $b$ = -23.5{\deg}).  
The reality of these two lanes is even more apparent in the IRAS 100 micron 
emission.  A third group of clouds lies almost directly below the Taurus 
clouds at an even more negative Galactic latitude (-45{\deg}); these very high-latitude 
clouds have a complicated velocity structure, with fragmented emission 
extending from -12 and +8 km s\(^{-1}\).  

As Figure 2 shows, this third group lies near 
the top of a rather prominent H I worm-like feature that appears to project up 
from the plane through a gap in the Taurus clouds near $l =$ 178{\deg}.  The 
gray levels in Figure 2 have been adjusted to emphasize this large 
H I feature, but it is discernable even in the all-sky, all-velocity 21 cm
map of Hartmann \& Burton (1997; p. 53). The molecular lanes just mentioned 
seem to project rather symmetrically to either side of the H I feature. 
The complicated velocity structure of the highest-latitude clouds and their
location near the top of the vertical H I structure suggest that they
may not have formed there, but rather may be fragments 
of the Taurus clouds that were expelled to higher latitudes, along with even 
more atomic gas, by events close to the plane.

\end{document}